\documentclass[12pt,twocolumn,preprintnumbers,amsmath,amssymb,fleqn]{revtex4}
\usepackage{graphics}
\normalsize

\begin{document}

\title{Optimal entanglement generation in cavity QED with dissipation}

\author{Jian-Song Zhang$^{1,2}$} \author{Jing-Bo Xu$^{1}$}%
\email{xujb@zju.edu.cn} \affiliation{$^{1}$ Zhejiang Institute of
Modern Physics and Physics Department, Zhejiang University, Hangzhou
310027, People's Republic of China \\
 $^{2}$ Department of Applied Physics, East China
Jiaotong University,  Nanchang 330013, People's Republic of China}

\begin{abstract}
{\bf Abstract}

 We investigate a two-level atom coupled to a cavity
with a strong classical driving field in a dissipative environment
and find an analytical expression of the time evolution density
matrix for the system. The analytical density operator is then used
to study the entanglement between the atom and cavity by considering
the competing process between the atom-field interactions and the
field-environment interactions. It is shown that there is an optimal
interaction time for generating atom-cavity entanglement.

\end{abstract}
\maketitle

\section{Introduction}
Quantum entanglement is one of the most striking features of quantum
mechanics\cite{Cirac2000, Nielsen2000}. It is central to many active
research fields, such as quantum computation\cite{Bennett2000},
quantum teleportation\cite{Bennett1993}, quantum
communication\cite{Gisin2007}, and quantum key
distribution\cite{Ekert1991}. The generation of quantum entangled
states has attracted much attention and many physical systems
including cavity quantum electrodynamics (Cavity
QED)\cite{Rauschenbeutel1999, Yonac2008, Bina2008, Li2005}, trapped
ions\cite{Cirac1995}, nuclear magnetic
resonance\cite{Gerhenfeld1997} and quantum dots\cite{Barenco1995},
have been suggested to generate entangled states.

In Ref.\cite{Solano2003}, Solano \emph{et al}. proposed a scheme to
generate the Schr\"{o}dinger cat states using two-level atoms
interacting with a cavity with the help of a strong classical
driving field. However, a quantum system is unavoidably influenced
by its surrounding environment\cite{Gardiner1991, Paz2002,
Angelo2006, Abdel2008}. The interaction between the quantum system
and its environment leads to decoherence which is the main problem
in the generation of quantum entanglement. In\cite{Lougovski2004},
the authors investigated the nonclassical properties of a cavity
field when no atom crosses it under the influence of dissipation by
making use of phase-space techniques.

In the present paper, we consider a quantum system consisting of a
two-level atom coupled to a cavity of high quality factor with a
strong classical driving field in a dissipative environment and find
an analytical expression of the time evolution density matrix for
the system. The analytical density matrix is then used to study the
entanglement and the purity of the system. In this model, there are
two competing processes in this system. One is the amplification
process introduced by the atom-field interactions (which leads to
entanglement). The other is the dissipation process due to the
field-environment interactions( which leads to disentanglment). Our
calculation shows that there is an optimal interaction time for
generating atom-cavity entanglement when the coupling constants and
the frequencies of the two-level atom, the cavity, and the classical
driving field are given. In section II, we solve the master equation
of the system by making use of the superoperator algebraic method
and obtain an explicit expression of time evolution density matrix
for the system of a two-level atom coupled to a cavity with a strong
classical driving field in a dissipative environment. In section
III, we use the concurrence to investigate the entanglement between
the two-level atom and the cavity by means of the analytical
expression of the density matrix for the system. In section IV, we
calculate the purity of the system by employing the linear entropy.
A conclusion is given in section V.

\section{Solution of an atom in a decay cavity with a strong classical driving field}
Now, we consider the time evolution of a two-level atom driven by an
external classical field in a dissipative cavity. The total
Hamiltonian of the system reads \cite{Solano2003}
\begin{eqnarray}
H&=&\omega a^{\dag}a+\frac{\omega_0}{2}\sigma_z+g_0(\sigma_+a
+\sigma_-a^{\dag})\nonumber\\
&&+\lambda_0(\sigma_+e^{-i\omega_c t}+\sigma_-e^{i\omega_c t}),
\end{eqnarray}
where $\omega$, $\omega_0$ and $\omega_c$ are the frequencies of the
cavity, atom and classical field, respectively. The operators
$\sigma_z$ and
 $\sigma_{\pm}$ are defined by  $\sigma_{z}=|0\rangle\langle 0|-|1\rangle\langle
 1|$, $\sigma_+=|0\rangle\langle 1|$, and $\sigma_-=|1\rangle\langle 0|$
where $|0\rangle$ and $|1\rangle$ are the excited and ground states
of the atom, respectively. Here, $a$ and $a^{\dag}$ are the
annihilation and creation operators of the cavity; $g_0$ and
$\lambda_0$ are the coupling constants of the interactions of the
atom with the cavity and with the classical driving field,
respectively. Note that we have set $\hbar=1$ throughout this paper.
For the sake of simplicity, we consider the resonant interaction
between one mode of the cavity and the two-level atom strongly
driven by the classical external field. The schematic picture of the
present system may be modeled as shown in Fig.1.

\begin{figure}
\centering {\rotatebox{360}{\includegraphics{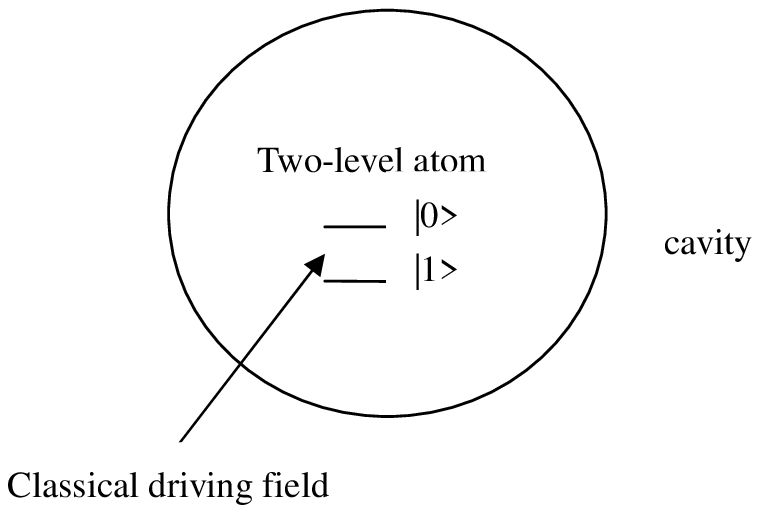}}}
 \caption{ The schematic picture of a two-level atom driven by an
external classical field in a dissipative cavity.
 }
\end{figure}

In the strong driving regime $\lambda_0\gg g_0$, we can realize a
rotating-wave approximation and eliminate from Eq.(1) the terms that
oscillate rapidly. The Hmailtonian of the system in the interaction
picture is \cite{Solano2003}
\begin{eqnarray}
V=g(a+a^{\dag})(\sigma_++\sigma_-),
\end{eqnarray}
with $g=g_0/2$. We assume that there is a reservoir coupled to the
electromagnetic field in the cavity. Thus, the master equation that
governs the dynamics of the system is\cite{Gardiner1991, Paz2002,
Lougovski2004, Scully1997}
\begin{eqnarray}
\frac{d\rho}{dt}&=&-i[V,\rho]+\mathcal{D}\rho
\nonumber\\
&=&-i[V,\rho]+k(2a\rho a^{\dag}-a^{\dag}a\rho-\rho
a^{\dag}a),
\end{eqnarray}
where $k$ is the decay constant and the superoperator
$\mathcal{D}=k(2a\cdot a^{\dag}-a^{\dag}a\cdot-\cdot a^{\dag}a)$
represents the losses in the cavity.

Now, we show how to obtain an analytical solution of the master
equation (3) with the help of superoperator algebraic method
\cite{Xu1999, Peixoto2004}. The density matrix for the system can be
represented as follows
\begin{eqnarray}
\rho(t)&=&\rho_{00}(t)\otimes|0\rangle\langle0|+\rho_{11}(t)\otimes|1\rangle\langle1|\nonumber\\
&&+\rho_{01}(t)\otimes|0\rangle\langle1|+\rho_{10}(t)\otimes|1\rangle\langle0|,
\end{eqnarray}
where $\rho_{ij}$'s are defined as $\rho_{ij}=\langle
i|\rho|j\rangle$, $i,j=0,1$. A straightforward calculation shows
that
\begin{eqnarray}
\dot{\rho}_{00}=-i g\{(a+a^{\dag})\rho_{10}-\rho_{01}(a+a^{\dag})\}+\mathcal{D}\rho_{00},\nonumber\\
\dot{\rho}_{01}=-i g\{(a+a^{\dag})\rho_{11}-\rho_{00}(a+a^{\dag})\}+\mathcal{D}\rho_{01},\nonumber\\
\dot{\rho}_{10}=-i g\{(a+a^{\dag})\rho_{00}-\rho_{11}(a+a^{\dag})\}+\mathcal{D}\rho_{10},\nonumber\\
\dot{\rho}_{11}=-i
g\{(a+a^{\dag})\rho_{01}-\rho_{10}(a+a^{\dag})\}+\mathcal{D}\rho_{11}.\nonumber\\
\end{eqnarray}
It can be proved that the above equations can be recast as
\begin{eqnarray}
\dot{\eta}_{\pm}(t)&=&\mathcal{L}_{\pm}\eta_{\pm}(t),\\
\dot{\eta'}_{\pm}(t)&=&\mathcal{L'}_{\pm}\eta'_{\pm}(t),
\end{eqnarray}
with
\begin{eqnarray}
\mathcal{L}_{\pm}&=&\pm i g(a\cdot+a^{\dag}\cdot-\cdot a-\cdot
a^{\dag})+\mathcal{D},\nonumber\\
\mathcal{L'}_{\pm}&=&\pm i g(a\cdot+a^{\dag}\cdot+\cdot a+\cdot
a^{\dag})+\mathcal{D},\nonumber\\
\eta_{\pm}&=&(\rho_{00}+\rho_{11})\mp(\rho_{01}+\rho_{10}),\nonumber\\
\eta'_{\pm}&=&(\rho_{00}-\rho_{11})\mp(\rho_{10}-\rho_{01}).
\end{eqnarray}
Here, the superoperators $a\cdot, \cdot a, a^{\dag}\cdot$ and $\cdot
a^{\dag}$ represent the action of creation and annihilation
operators on an operator
\begin{eqnarray}
(a\cdot)o&=&ao, (\cdot a)o=oa,\nonumber\\
(a^{\dag}\cdot)o&=&a^{\dag}o, (\cdot a^{\dag})o=oa^{\dag}.
\end{eqnarray}
Combing the communication relation $[a,a^{\dag}]=1$ with the above
equation, we derive the following communication relations between
the superoperators
\begin{eqnarray}
[(a\cdot),(a^{\dag}\cdot)]=1, \quad [(\cdot a),(\cdot a^{\dag})]
=-1,
\end{eqnarray}
while all other communication relations are zero. It is easy to
check that the superoperators  $\mathcal{M}=a\cdot a^{\dag}$,
$\mathcal{R}=a^{\dag}a\cdot$ and $\mathcal{L}=\cdot a^{\dag}a$
satisfy the relations
\begin{eqnarray}
&&[\mathcal{R},\mathcal{M}]=[\mathcal{L},\mathcal{M}]=-\mathcal{M},
[\mathcal{R},\mathcal{L}]=0.
\end{eqnarray}

We assume the field is initially in coherent state $|\alpha\rangle$
and the atom is in the exited state $|0\rangle$. Thus the initial
state of the atom-field system is
\begin{eqnarray}
|\psi(0)\rangle=|\alpha\rangle\otimes|0\rangle.
\end{eqnarray}
Then the elements of the density matrix are initially
$\rho_{00}(0)=|\alpha\rangle\langle\alpha|$ and
$\rho_{11}(0)=\rho_{01}(0)=\rho_{10}(0)=0$ which leads to
$\eta_{\pm}(0)=\eta_{\pm}'(0)=|\alpha\rangle\langle\alpha|$.

Using the above results we find the solution of the master equations
(6) by making use of the superoperator algebraic method
\cite{Xu1999}
\begin{eqnarray}
\eta_{\pm}(t)&=&e^{\mathcal{L}_{\pm}t}\eta_{\pm}(0)\nonumber\\
&=&e^{\pm i gt
\mathcal{Y}_1+kt\mathcal{X}_1}(|\alpha\rangle\langle\alpha|)\nonumber\\
&=&e^{kt\mathcal{X}_1}e^{\frac{\mp i g}{k}(1-e^{k
t})\mathcal{Y}_1}(|\alpha\rangle\langle\alpha|)\nonumber\\
&=&e^{(e^{2kt}-1)\mathcal{M}}e^{-kt\mathcal{R}}e^{-kt\mathcal{L}}
\nonumber\\
&&\times e^{\frac{\mp i g}{k}(1-e^{k
t})\mathcal{Y}_1}(|\alpha\rangle\langle\alpha|)\nonumber\\
&=&|\alpha_{\pm}(t)\rangle\langle\alpha_{\pm}(t)|,
\end{eqnarray}
where
\begin{eqnarray}
|\alpha_{\pm}(t)\rangle&=&|\alpha e^{-kt}\pm\frac{i
g}{k}(1-e^{-kt})\rangle, \nonumber\\
\mathcal{X}_1&=&2\mathcal{M}-\mathcal{R}-\mathcal{L}, \nonumber\\
\mathcal{Y}_1&=&a\cdot+a^{\dag}\cdot-\cdot a-\cdot a^{\dag}.
\end{eqnarray}

Similarly, the solutions of the master equations (7) can also be
derived as
\begin{eqnarray}
\eta'_{\pm}(t)&=&e^{\pm i g t(\mathcal{X}_2+\mathcal{Y}_2)+k t
\mathcal{X}_1}|\alpha\rangle\langle\alpha|\nonumber\\
&=&e^{f_1}e^{f_2\mathcal{Y}_2}e^{kt\mathcal{X}_1}e^{f_2
\mathcal{X}_2}|\alpha\rangle\langle\alpha|,\nonumber\\
&=&f_{\pm}(t)|\alpha_{\pm}(t)\rangle\langle\alpha_{\mp}(t)|,
\end{eqnarray}
where
\begin{eqnarray}
\mathcal{X}_2&=&2(a\cdot+\cdot a^{\dag} ),\nonumber\\
\mathcal{Y}_2&=&-a\cdot+a^{\dag}\cdot-\cdot a^{\dag}+\cdot a, \nonumber\\
f_1&=&\frac{-4g^2}{k^2}(e^{-k t}-1+k t), \nonumber\\
f_2&=&\frac{i g}{k}(1-e^{-k t}), \nonumber\\
f_{\pm}&=&e^{f_1+2|f_2|^2\pm f_2(\alpha+\alpha^*)(2-e^{-kt})},
\end{eqnarray}

Combing the solutions of the Eqs.(6) and (7), we obtain the explicit
expression of the density matrix of the system at time t
\begin{eqnarray}
\rho(t)&=&\rho_{00}(t)\otimes|0\rangle\langle0|+\rho_{11}(t)\otimes|1\rangle\langle1|\nonumber\\
&&+\rho_{01}(t)\otimes|0\rangle\langle1|+\rho_{10}(t)\otimes|1\rangle\langle0|,
\end{eqnarray}
with
\begin{eqnarray}
\rho_{00}(t)&=&\frac{1}{4}[|\alpha_+(t)\rangle\langle\alpha_+(t)|+|\alpha_-(t)\rangle\langle\alpha_-(t)|\nonumber\\
&&
+f_+(t)|\alpha_+(t)\rangle\langle\alpha_-(t)|\nonumber\\
&&+f_-(t)|\alpha_-(t)\rangle\langle\alpha_+(t)|],\nonumber\\
\rho_{11}(t)&=&\frac{1}{4}[|\alpha_+(t)\rangle\langle\alpha_+(t)|+|\alpha_-(t)\rangle\langle\alpha_-(t)|\nonumber\\
&&
-f_+(t)|\alpha_+(t)\rangle\langle\alpha_-(t)|\nonumber\\
&&-f_-(t)|\alpha_-(t)\rangle\langle\alpha_+(t)|],\nonumber\\
\rho_{01}(t)&=&-\frac{1}{4}[|\alpha_+(t)\rangle\langle\alpha_+(t)|-|\alpha_-(t)\rangle\langle\alpha_-(t)|\nonumber\\
&&
-f_+(t)|\alpha_+(t)\rangle\langle\alpha_-(t)|\nonumber\\
&&+f_-(t)|\alpha_-(t)\rangle\langle\alpha_+(t)|],\nonumber\\
\rho_{10}(t)&=&\rho^{\dag}_{01}(t).
\end{eqnarray}

Thus, we find an analytical solution of the master equation with the
initial state $|\psi(0)\rangle=|\alpha\rangle\otimes|0\rangle$. In
the following sections, the explicit expression of density matrix is
used to investigate the entanglement and the purity of the system.

\section{Entanglement between a two-level atom and decay cavity }
It is well known that entangled states are the basic resource of
quantum information processing, such as quantum communication and
quantum teleportation. Cavity QED is a useful tool to generate
entangled states. However, an entangled state will become mixed
and/or less entangled due to the decay of cavities. Here, we confine
our consideration in the question whether we can choose an optimal
combination of parameters $g$, $k$, and $t$ to maximize the
entanglement of the atom-field system. In order to study the
entanglement of above system described by density matrix $\rho$, we
adopt the measure concurrence which is defined as\cite{Wootters1998}
\begin{equation}
C=\max{\{0, \lambda_1-\lambda_2-\lambda_3-\lambda_4\}},
\end{equation}
where the $\lambda_i$(i=1,2,3,4) are the square roots of the
eigenvalues in decreasing order of the magnitude of the
``spin-flipped" density matrix operator
$R=\rho(\sigma_y\otimes\sigma_y)\rho^*(\sigma_y\otimes\sigma_y)$ and
$\sigma_y$ is the Pauli Y matrix, i.e., $\sigma_y= \left(
\begin{array}{cc}
  0 & -i \\
  i & 0
  \end{array}
  \right)$.

In general, the model consisting of the two-level atom and the field
is a $2\times\infty$ system. However, for the initial state
$|\psi(0)\rangle=|\alpha\rangle\otimes|0\rangle$, the atom-field
system can be mapped onto a $2\times2$ system as one can see from
the density matrix. To this end we introduce two orthonormal vectors
$|\uparrow\rangle$ and $|\downarrow\rangle$ which are defined by
\begin{eqnarray}
|\uparrow\rangle&=&|\alpha_+(t)\rangle,\\
|\downarrow\rangle&=&\frac{1}{\sqrt{1-|\lambda(t)|^2}}(|\alpha_-(t)\rangle-\lambda(t)|\alpha_+(t)\rangle),\nonumber
\end{eqnarray}
with $\lambda(t)=\langle\alpha_+(t)|\alpha_-(t)\rangle$. Therefor,
the states $|\alpha_+(t)\rangle$ and $|\alpha_-(t)\rangle$ can be
represented in terms of $|\uparrow\rangle$ and $|\downarrow\rangle$
\begin{eqnarray}
|\alpha_+(t)\rangle&=&|\uparrow\rangle,\nonumber\\
|\alpha_-(t)\rangle&=&\lambda(t)|\uparrow\rangle+\sqrt{1-|\lambda(t)|^2}|\downarrow\rangle.
\end{eqnarray}
Inserting Eq.(21) into Eq.(18), one can rewrite the state of the
atom-field system at time t, i.e., the state of system can be mapped
onto a system composed by two two-dimensional subsystems. Note that
the density matrix of Eq.(17) is now an effective two-qubit system
whose entanglement can be evaluated by the measure concurrence. With
the help of Eqs.(17), (18), (19), and (21), one can obtain the
concurrence the density matrix of Eq.(17) which is too long to
present here. In Fig.2, we plot the entanglement dynamics of the
quantum system as a function of time t. It is worth noting that we
have used the dimensionless quantities throughout this paper.

We now turn to show that our model can be used to generate the
Schr\"{o}dinger cat state in the case of $\alpha=k=0$. A simple
calculation shows that
\begin{eqnarray}
\alpha_{\pm}(t)&=&\pm i g t, f_1(t)=-2g^2t^2, \nonumber\\
 f_2(t)&=&i g
t,
f_{\pm}=1,\nonumber\\
\eta_{\pm}(t)&=&|\pm i g t\rangle\langle \pm i g t|,\nonumber\\
\eta'_{\pm}(t)&=&|\pm i g t\rangle\langle \mp i g t|.
\end{eqnarray}
Here, we have used the fact $\alpha=k=0$. Thus the atom-field state
at time t is
\begin{eqnarray}
|\psi(t)\rangle&=&\frac{1}{2}[(|-i g t\rangle+|i g
t\rangle)\otimes|0\rangle\nonumber\\
&&+(|-i g t\rangle-|i g
t\rangle)\otimes|1\rangle]\nonumber\\
&=&\frac{1}{\sqrt{2}}(|-i g t\rangle|+\rangle-|i g
t\rangle|-\rangle),
\end{eqnarray}
with
\begin{eqnarray}
|\pm\rangle=\frac{1}{\sqrt{2}}|1\rangle\pm|0\rangle.
\end{eqnarray}
The entangled state $|\psi(t)\rangle$ is the Schr\"{o}dinger cat
state.

In the case of $k>0$, entangled states will become mixed and/or less
entangled due to the decay of cavities. Fortunately, one can find
that there is an optimal time $t_{opt}$ to maximize the entanglement
if the coupling constant g and the decay rate of the cavity k are
given. In Fig.2, we plot the concurrence of the atom-field system as
a function of time t and the coupling constant g for $k=0$ (Upper
panel) and $k=0.05$ (Lower panel). It is not difficult to see that
in the case of $k=0$ the concurrence first increases with time t and
then reaches a plateau. However, when $k>0$ there is no plateau for
the concurrence. It first increases with time t, and reaches the
maximal value, then decreases with time t. Finally, the atom and the
field is disentangled. In order to see this more clearly, we plot
the concurrence as a function of time for different values of decay
constant k in Fig.3. One can easily find that if the coupling
constant g and the decay rate of the cavity k are known one can find
an optimal time $t_{opt}$ to maximize the concurrence. For instance,
the optimal time $t_{opt}$ are 0.83 and 0.61 in the case of $g=1,
k=0.1$ and $g=1, k=1$, respectively.

Physically, this is a result of two competing processes. One is the
amplification process due to the interaction between the atom and
the cavity. The other is the dissipation process due to the
field-environment interaction. These two processes compete against
each other as the system evolves. At first, the amplification
process dominates the dissipation process and the entanglement of
the system increases with time. Later, the dissipation process is in
control and the entanglement decreases with time. Finally, the
atom-field system is disentangled as one can see from Fig.3.  One
can also use the negativity to evaluate the degree of entanglement
of the two-qubit system ( e.g., see Ref.\cite{Vidal2002}). For a
$2\times2$ quantum system the concurrence and negativity are both
good entanglement measures. In fact it has been proved that the
negativity of a state is always lager than
$\sqrt{(1-C)^2+C^2}-(1-C)$ and smaller than C, where C is the
concurrence of a state\cite{Ver2001}. For a $2\times3$ quantum
system the concurrence is not applicable while the negativity is
still valid\cite{Vidal2002}.

\begin{figure}
\centering {\includegraphics{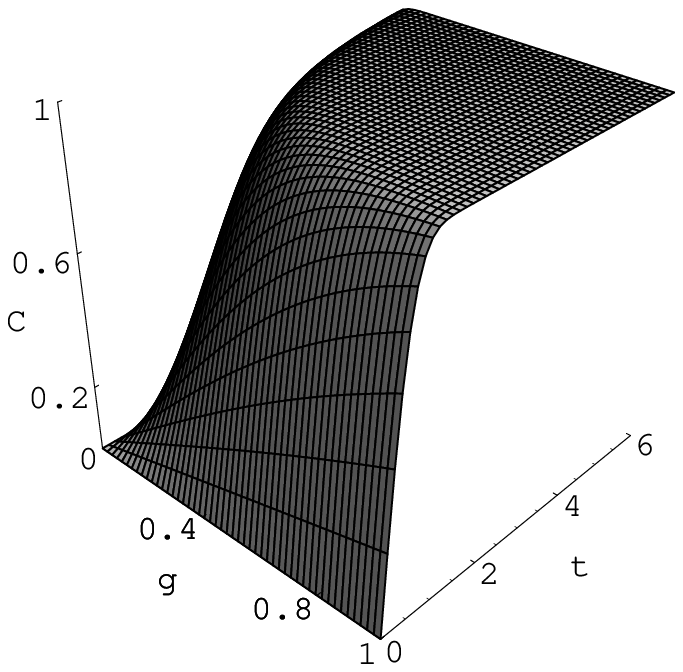}}\\
\centering{\includegraphics{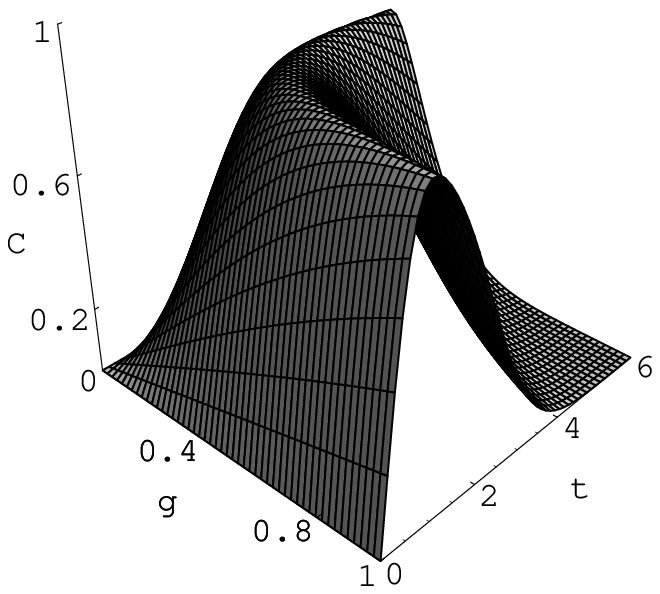}}
 \caption{ The concurrence is plotted as a function of g
and t with $\alpha=1$. Upper panel: The decay rate constant k is
zero. Lowe panel: The decay constant k is 0.0.5.
 }
\end{figure}

\begin{figure}
\centering {\rotatebox{360}{\includegraphics{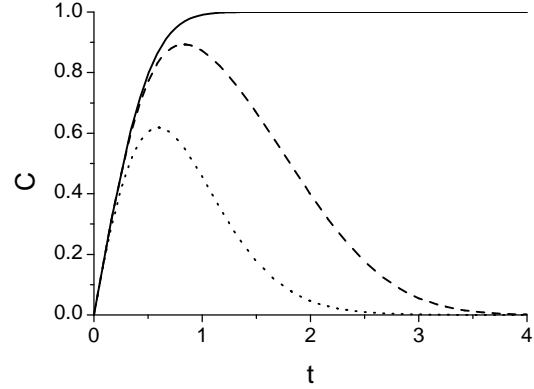}}}
 \caption{ The concurrence is plotted as a function of t with $g=1$ for $k=0$(solid line),
 $k=0.1$(dash line) and $k=1$(dot line).
 }
\end{figure}

\section{Purity of system}
In this section, we investigate the purity of the system by
employing the linear entropy. Many protocols in quantum information
processing require pure, maximally entangled quantum states. For
example, quantum teleportation often relies heavily on the purity
and entanglement of the initial state. However,  an pure and
entangled quantum system usually becomes mixed and/or less entangled
under the influence of decoherence. Here, we adopt the linear
entropy to quantify the mixedness of a state defined by
\begin{eqnarray}
S(\rho)=1-Tr(\rho^2).
\end{eqnarray}
Generally, if $\rho$ is the density matrix of a pure state, $s=0$,
otherwise $s>0$. It has also been proved that a bipartite mixed
states is useless for quantum teleportation if its linear entropy
exceeds $1/2$ for a two qubits system. The purity of the atom-field
system is\cite{Peixoto2004}
\begin{eqnarray}
S(\rho)&=&1-Tr(\rho^2)=1-Tr_F\{Tr_A(\rho^2)\}\nonumber\\
&=&1-Tr_F\{(\rho_{00}+\rho_{11})^2\}\nonumber\\
&=&\frac{1}{2}(1-|f_+(t)|^2),
\end{eqnarray}
where we have used the fact
\begin{eqnarray}
Tr_F(|\alpha_+(t)\rangle\langle\alpha_-(t)|)=\langle\alpha_-(t)|\alpha_+(t)\rangle.
\end{eqnarray}

In order to analyze the purity of the two-level atom at time t, we
first trace out the field's variables and obtain
\begin{eqnarray}
\rho_A(t)&=&\frac{1}{2}\{1+Re[f_+(t)\lambda(t)]\}|\uparrow\rangle\langle\uparrow|\nonumber\\
&&+\frac{1}{2}\{1-Re[f_+(t)\lambda(t)]\}|\downarrow\rangle\langle\downarrow|\\
&&+\frac{i
Im[f_+(t)\lambda(t)]}{2}(|\uparrow\rangle\langle\downarrow|-|\downarrow\rangle\langle\uparrow|),\nonumber
\end{eqnarray}
with $Re$ and $Im$ denoting the real and the image part of a complex
number, respectively. The linear entropy of the atom is
\begin{eqnarray}
S_A(\rho_A)=\frac{1}{2}[1-|f_+(t)\lambda(t)|^2].
\end{eqnarray}

Finally, we investigate the purity of the field. After tracing out
the atom's variables, the density matrix of the field is
\begin{eqnarray}
\rho_F(t)&=&\rho_{00}(t)+\rho_{11}(t)\\
&=&\frac{1}{2}[|\alpha_+(t)\rangle\langle\alpha_+(t)|+|\alpha_-(t)\rangle\langle\alpha_-(t)|].\nonumber
\end{eqnarray}
One can write the linear entropy of the field as follows
\begin{eqnarray}
S_F(\rho_F)=\frac{1}{2}[1-|\lambda(t)|^2].
\end{eqnarray}
By using the explicit expression of the functions $|f_+(t)|^2$ and
$|\lambda(t)|^2$, we find that the linear entropies $S$, $S_A$, and
$S_F$ are independent of the parameter $\alpha$. The linear entropy
is plotted as a function of time t for several values of $k/g$ in
Figs.(4-6). From Fig.4, we can see that the linear entropies $S$,
$S_A$, and $S_F$ increase with time and then reach the maximal value
$0.5$. However, the situation is different when the parameter $k/g$
increases. In Fig.5 and Fig.6, it is easy to see that the maximal
values of $S_F$ (dotted line) is less than $0.5$. In general, the
linear entropy $S_A$ is larger than the linear entropy $S$. As the
parameter $k/g$ increases, the linear entropy $S_F$ decreases
significantly(see the dotted line in Fig.6). This result is similar
to that obtained in Ref.\cite{Peixoto2004}. The atomic and field
coherence loss can be measured by the linear entropy $S_A$ and $S_F$
\cite{Peixoto2004}. Comparing Figs.(4-6), one can see that the
atomic coherence $S_A$ and the field coherence $S_F$ decrease with
the increase of the decay rate k. However, the atom-field system
preserves its linear entropy $S$ as the system evolves. For example,
the linear entropy of the atom-field system is 0.5 if $t>5$(see the
solid lines in Figs.(4-6)) which is a result of two competing
processes. One is the process due to the interaction between the
atom and the field. The other is the process due to the
field-environment interaction. As we have pointed out, these two
processes compete against each other during the whole evolution. The
linear entropy $S$ first increases with time t and then reaches a
plateau. We note that the similar features have also been observed
in Ref.\cite{Abdel2008} where a Cooper-pair box is put into a
phase-damped cavity.

\begin{figure}
\centering {\rotatebox{360}{\includegraphics{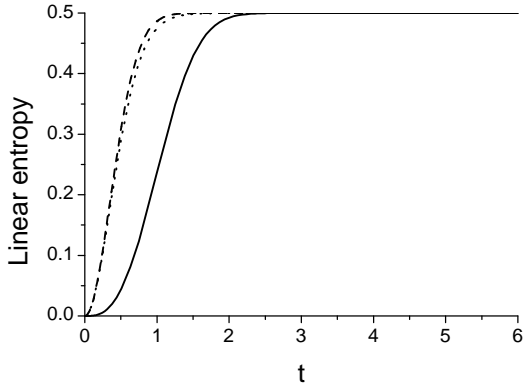}}}
 \caption{ The linear entropy is plotted as a function of t with $k/g=0.3$ for $S$(solid line),
 $S_F$(dotted line) and $S_A$(dashed line).
 }
\end{figure}

\begin{figure}
\centering {\rotatebox{360}{\includegraphics{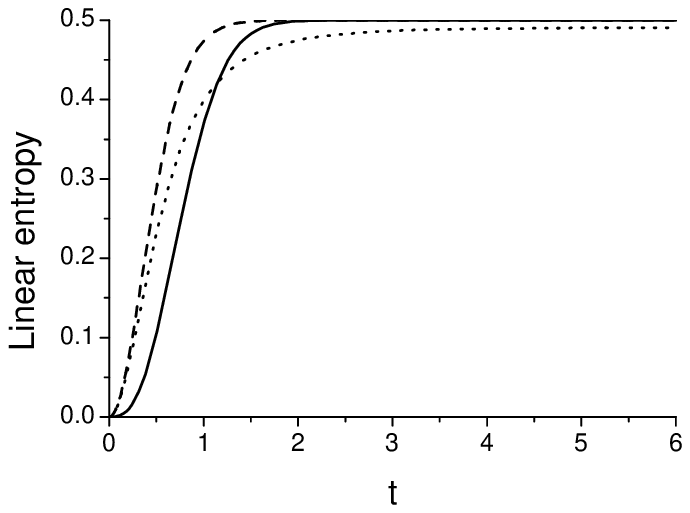}}}
 \caption{ The linear entropy is plotted as a function of t with $k/g=1$ for $S$(solid line),
 $S_F$(dotted line) and $S_A$(dashed line).
 }
\end{figure}

\begin{figure}
\centering {\rotatebox{360}{\includegraphics{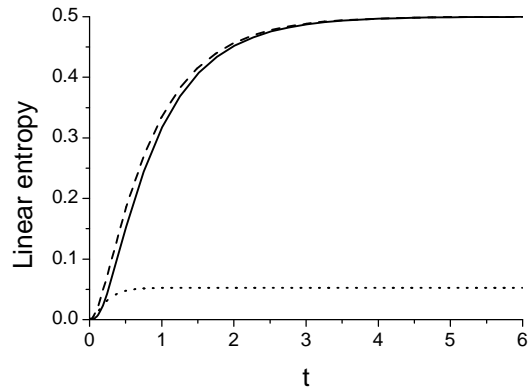}}}
 \caption{ The linear entropy is plotted as a function of t with $k/g=6$ for $S$(solid line),
 $S_F$(dotted line) and $S_A$(dashed line).
 }
\end{figure}

\section{Conclusions}
In the present paper, we have investigated entanglement dynamics of
a quantum system consisting of one two-level atom coupled to a
dissipative cavity with a strong classical driving field. An
analytical expression of the time evolution density matrix operator
for the system is found and used to study the entanglement and the
purity of the system. There are two competing processes. One is the
amplification process due to the interaction between the atom and
the cavity. The other is the dissipation process due to the
field-environment interaction. Our calculation shows that there is
an optimal interaction time to maximize the entanglement of the
atom-field system in the presence of dissipation when the coupling
constants and the frequencies of the two-level atom, the cavity, and
the classical driving field are given. The approach adopted in the
present paper may be extended to systems of two or more two-level
atoms in dissipative cavities.

\section*{Acknowledgments}
 This project was supported by the National Natural Science Foundation of
China(Grant no. 10774131) and the National Key Project for
Fundamental Research of China (Grant no. 2006CB921403).

\end{document}